\documentclass[aps,pra,twocolumn,showpacs,superscriptaddress,floatfix,nofootinbib]{revtex4}
\usepackage{graphicx}
\usepackage{amsmath}
\usepackage{epsfig}
\usepackage{helvet} 
\usepackage{amssymb}

\newcommand{\be}{\begin{equation}}
\newcommand{\ee}{\end{equation}}
\newcommand{\bea}{\begin{eqnarray}}
\newcommand{\eea}{\end{eqnarray}}

\newcommand{\tr}{\mbox{tr}}

\newcommand{\ket}[1]{\mbox{$| #1 \rangle$}}

\def\tr{ \mbox{tr}}

\begin{document}

\title{
Simulation of interacting fermions with entanglement renormalization}

\author{Philippe Corboz}
\affiliation{School of Mathematics and Physics, The University of
Queensland, Queensland 4072, Australia} 
\author{Glen Evenbly}
\affiliation{School of Mathematics and Physics, The University of
Queensland, Queensland 4072, Australia}
\author{Frank Verstraete}
\affiliation{Fakult${\it \ddot{{\rm {\it  a}}}}$t f${\it \ddot{{\rm {\it u}}}}$r Physik, Universit${\it \ddot{{\rm {\it  a}}}}$t Wien, Boltzmanngasse 3, A-1090 Wien, Austria}
\author{Guifr\'e Vidal} 
\affiliation{School of Mathematics and Physics, The University of
Queensland, Queensland 4072, Australia} 

\date{\today}

\begin{abstract}
We propose an algorithm to simulate interacting fermions on a two dimensional lattice. The approach is an extension of the entanglement renormalization technique [Phys. Rev. Lett. 99, 220405 (2007)] and the related multi-scale entanglement renormalization ansatz. Benchmark calculations for free and interacting fermions on lattices ranging from $6\times 6$ to $162\times 162$ sites with periodic boundary conditions confirm the validity of this proposal.
\end{abstract}


\maketitle

The simulation of interacting fermions is of capital importance for our understanding of strongly correlated phenomena such as high-temperature superconductivity or the fractional quantum Hall effect. Quantum Monte Carlo techniques, so successful in addressing bosonic many-body problems, fail for fermionic systems due to the so-called sign problem \cite{sign}. Many alternative techniques have been used, including exact diagonalization, density matrix renormalization group, dynamical cluster approximation, or variational and Gaussian Monte Carlo methods \cite{methods}. But despite all these approaches, even the ground state properties of basic lattice models for interacting fermions, such as the Hubbard model, remain highly controversial in two spatial dimensions. 

Recently, \emph{entanglement renormalization} \cite{ER} has been proposed to efficiently simulate quantum systems on a lattice. By means of a coarse-graining transformation, the size of the system is progressively reduced until exact diagonalization becomes feasible. The key of the approach is the use of \emph{disentanglers} to remove short-range entanglement at each coarse-graining step. In this way the low energy properties of the system are preserved while the computational cost is kept under control. An approximation to the ground state of the lattice is then encoded in the \emph{multi-scale entanglement renormalization ansatz} (MERA) \cite{MERA}, from which one can compute the expected value of local observables and correlators. The scheme is scalable and has been used to address arbitrarily large, two-dimensional spin systems \cite{2D}. 

In this paper we show that entanglement renormalization can be adapted to address fermionic systems. We first explain the key idea underlying the fermionic version of this approach and then demonstrate its validity by computing the ground state of fermionic systems on lattices of up to $162\times 162$ sites. As in the bosonic case, the cost of simulations does not depend on whether the particles interact, but rather on the amount of entaglement present in the ground state.

\textbf{Fermions on a lattice.---}
We consider a system of fermions on a lattice $\mathcal{L}$ made of $N$ sites, where each site $r\in \mathcal{L}$ is described by a single fermionic operator $c_r$, with
\begin{equation}
	\{c_r, c_{s}^{\dagger}\} = \delta_{rs}I_r,~~~~~\{c_r, c_{s}\} = 0.
	\label{eq:anti}
\end{equation}
[This simple setting is later generalized.]
The system is further characterized by a parity preserving Hamiltonian $H$ that decomposes as an even polynomial of fermionic operators (see e.g. Eqs. \ref{eq:free}-\ref{eq:Hint}), that is as a sum of even powers of the $c_r$'s, such as $c_r c_s$, $c_r c_s^\dagger$ or $c^{\dagger}_r c_r c^{\dagger}_s c_s$. The occupation number basis for the $2^N$-dimensional vector space of the lattice is given by
\begin{equation}
 \ket{i_1 i_2 \cdots i_N} \equiv (c_1^\dagger)^{i_1} (c_2^\dagger)^{i_2} \cdots(c_N^\dagger)^{i_N}\ket{00\cdots 0},
\end{equation}
where $i_r=0,1$ indicates the absence/presence of a fermion on site $r$ and the vacuum state $\ket{00\cdots 0}$ corresponds to not having any fermion in the lattice. With the usual Jordan-Wigner transformation \cite{sigma},
\begin{equation}
	c_r = (Z_1 \cdots Z_{r-1}) \sigma_r, ~~~Z_{r} \equiv I_r-2c_r^{\dagger} c_r,
\end{equation}
fermionic operators are mapped into spin operators $\sigma_r$,
\begin{equation}
	[\sigma_r,\sigma_{s}^{\dagger}] = \delta_{rs}Z_r,~~~~~[\sigma_r, \sigma_{s}]= 0,
\end{equation}
where $Z_r$ acts diagonally on the occupation number basis of site $r$, $Z_r\ket{0_r} = \ket{0_r}$ and $Z_r\ket{1_r} = - \ket{1_r}$, and thus preserves the parity of the occupation number of a state, whereas the spin operator $\sigma_r$ changes parity, $\sigma_r\ket{1_r} = \ket{0_r}$, $\sigma_r\ket{0_r} = 0$. When expressed in spin variables, some operators, such as $c^{\dagger}_r c_r c^{\dagger}_s c_s=\sigma^{\dagger}_r \sigma_r \sigma^{\dagger}_s \sigma_s$, remain supported on just the same sites. Instead, other operators, such as $c^{\dagger}_r c_s$, develop a string of $Z$'s,
\begin{eqnarray}
	c^{\dagger}_r c_s = \sigma^{\dagger}_r (Z_{r+1} \cdots Z_{s-1})\sigma_s.
\label{eq:cc}
\end{eqnarray}
In the latter case we say that the \emph{bosonic support} of the operator is larger than its \emph{fermionic support}.

\begin{figure}[!htb]
\begin{center}
\includegraphics[width=7cm]{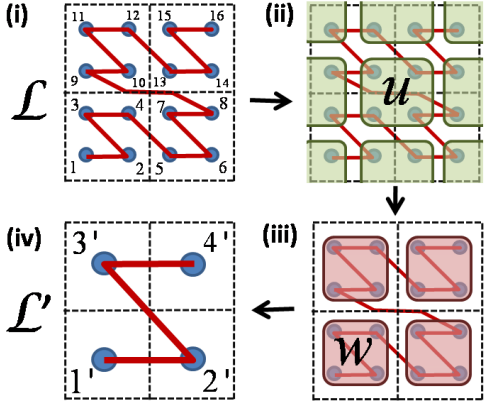}
\caption{
(Color online) (i) A $4\times4$ lattice $\mathcal{L}$ is coarse-grained by first (ii) applying disentanglers $u$ on blocks of four sites and then (iii) isometries $w$ also on blocks of four sites, producing (iv) a $2\times 2$ lattice $\mathcal{L}'$. Notice that the Jordan-Wigner order for the sites is such that each isometry $w$ acts on four consecutive sites. 
}
\label{fig:ER}
\end{center}
\end{figure}

\textbf{Coarse-graining transformation.---}
Following the formalism of entanglement renormalization \cite{ER,MERA}, our goal is to coarse-grain the lattice $\mathcal{L}$ into a smaller lattice $\mathcal{L}'$. This is achieved in two steps, as exemplified in Fig. \ref{fig:ER}(i-iv). First, disentanglers $u$ are applied across the boundary of blocks of sites of $\mathcal{L}$.
Second, isometries $w$ are used to map each block of sites of $\mathcal{L}$ into a single site of the coarse-grained lattice $\mathcal{L}'$. As extensively discussed in Refs. \cite{ER,MERA,2D,algorithm}, a fundamental aspect of entanglement renormalization, on which the efficiency of the approach rests, is that local operators remain local under successive iterations of the coarse-graining. 

Here we shall argue that, with a proper choice of disentanglers and isometries, locality of operators is preserved also in the fermionic case. Two main difficulties need to be addressed. On the one hand, fermionic operators such as the hoping term $c^{\dagger}_r c_s$ in Eq. \ref{eq:cc}, have non-local bosonic support, with a string of $Z$'s that may involve up to $O(N)$ lattice sites. Such non-local supports could in principle preclude the use of entanglement renormalization, an approach based on transforming local operators only. It turns out, however, that this difficulty can be circumvented by focusing on the fermionic support of such operators, which is local. On the other hand, the locality of fermionic supports is not preserved by the type of disentanglers and isometries used to coarse-grain bosonic systems. This can be understood by observing that such disentanglers and isometries are themselves bosonic (commuting) operators, and bosonic and fermionic (anticommuting) operators are mutually non-local. Fortunatelly, this difficulty can be resolved by replacing \emph{bosonic} disentanglers $u$ and isometries $w$ with \emph{fermionic} counterparts, that is, with tensors $u$ and $w$ that are local when written in fermionic variables. This is further illustrated next with an explicit example by considering the simple $4\times 4$ square lattice of Fig. \ref{fig:ER}.

\begin{figure}[!htb]
\begin{center}
\includegraphics[width=7.5cm]{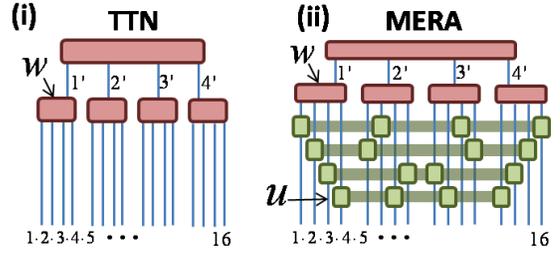}
\caption{
(Color online) (i) TTN to approximate the ground state $\ket{\Psi}$ of $H$ obtained by adding a top tensor (representing the state of $\mathcal{L}'$) to the isometries $w$ of Fig. \ref{fig:ER}. Notice that the sites (vertical lines) are arranged in 1D according to the Jordan-Wigner order. Isometries $w$ are local even when written in terms of spin operators. (ii) MERA obtained by adding disentanglers $u$ to the TTN according to Fig. \ref{fig:ER}. Disentanglers $u$ are delocalized and decompose into sums of terms that contain strings of $Z$'s (represented with ribbons).} 
\label{fig:MERA}
\end{center}
\end{figure}

\begin{figure}[!ht]
\begin{center}
\includegraphics[width=8cm]{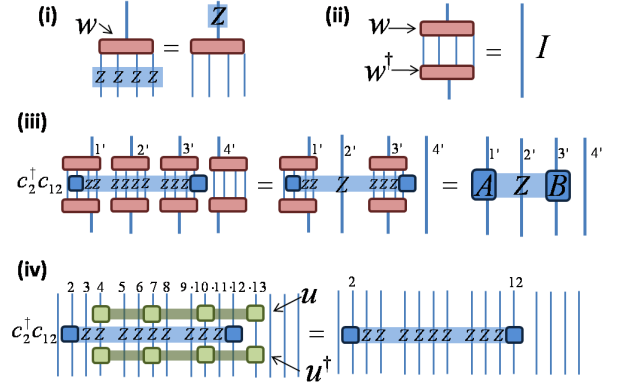}
\caption{
(Color online) (i) A fermionic isometry $w$ preserves parity, cf. Eq. \ref{eq:parity}. (ii) By construction, an isometry $w$ is such that the pair $w$ $w^{\dagger}$ anihilates into the identity $I$. (iii) Coarse-graining of the operator $c_{2}^{\dagger}c_{12}$ into $A_{1'}Z_{2'}B_{3'}$, cf. Eq. \ref{eq:coarse}, by using properties (i) and (ii). (iv) The operator $c_{2}^{\dagger}c_{12}$ also commutes with a disentangler $u$ if $c_{2}^{\dagger}c_{12}$ and $u$ have disjoint fermionic support.} 
\label{fig:Zs}
\end{center}
\end{figure}

\textbf{Fermionic isometries.---} 
Let us first assume that the disentanglers in Fig. \ref{fig:ER}($ii$) are the identity operator, $u=I$, that is, lattice $\mathcal{L}'$ is obtained from $\mathcal{L}$ by the use of isometries $w$ that map four sites into one. 
In this case the isometries, together with a top tensor, form an ansatz for the ground state $\ket{\Psi}$ of $H$ known as \emph{tree tensor network} (TTN) \cite{TTN}, see Fig. \ref{fig:MERA}($i$). 
%
%
Notice that each isometry $w$ is parity preserving (i.e. built as an even polynomial of $c$ and $c^{\dagger}$ operators) that act on a block of four \emph{consecutive} sites, e.g. sites $1,2,3,4 \in \mathcal{L}$. Therefore, the fermionic and bosonic supports of $w$ coincide (that is, when $w$ is expressed in terms of spin operators, there are no strings of $Z$'s leaving the block). If we assume (temporarily) that the coarse-grained site $1'\in\mathcal{L}'$ is also described by a two-dimensional space with corresponding operator $Z_{1'}$, then the parity symmetry of $w$ implies 
\begin{eqnarray}
w = (Z_1 Z_2  Z_3  Z_4) w Z_{1'},
\label{eq:parity}
\end{eqnarray}
and strings of $Z$'s simply 'commute' with the coarse-graining transformation, see Fig. \ref{fig:Zs}($i$).

It follows that, under coarse-graining, an operator with local fermionic support, say $c^{\dagger}_2c_{12}$, is transformed into an operator whose fermionic support is also local,
\begin{equation}
	\sigma^{\dagger}_2 \left(Z_3 Z_4 \cdots Z_{11} \right) \sigma_{12} \rightarrow A_{1'} Z_{2'} B_{3'},
	\label{eq:coarse}
\end{equation}
where $A_{1'}$ and $B_{3'}$ are parity-changing operators, a property they inherit from $\sigma_2^{\dagger}$ and $\sigma_{12}$, see Fig. \ref{fig:Zs} (iii). Importantly, $A_{1'}$ and $B_{3'}$ are obtained by coarse-graining operators $\sigma_2^{\dagger}$ and $\sigma_{12}$ respectively, whereas the original string of $Z$'s simply shrinks into $Z_{2'}$. In other words, in spite of the presence of a string of $Z$'s, all the manipulations involved in the coarse-graining of $c^{\dagger}_2c_{12}$ by fermionic isometries $w$ can be performed locally and thus efficiently.

\textbf{Fermionic disentanglers.---}
Let us now consider non-trivial fermionic disentanglers $u\neq I$, which together with the isometries $w$ and a top tensor constitute the MERA for the ground state $\ket{\Psi}$ of $H$, see Fig. \ref{fig:MERA}($ii$). Again, fermionic disentanglers are built as even polynomials of $c$ and $c^{\dagger}$ operators, but since they are not supported on consecutive sites of $\mathcal{L}$, they include strings of $Z$'s when written in terms of spin variables. 

One may fear that such strings of $Z$'s may turn local operators into highly non-local ones. However, this is not the case for operators with a local fermionic support. As it can be easily checked from Eq. \ref{eq:anti}, two even polynomials of $c$ and $c^{\dagger}$ operators with disjoint fermionic support commute with each other. That means, for instance, that the operator $c_2^{\dagger}c_{12}$ in Eq. \ref{eq:coarse} commutes with a disentangler $u$ with fermionic support on sites $4,7,10,13\in \mathcal{L}$, $u(c_2^{\dagger}c_{12})u^{\dagger} = c_2^{\dagger}c_{12}$, see Fig. \ref{fig:Zs}($iv$).  
That is, fermionic disentanglers only expand the fermionic support of operators as much as bosonic disentanglers do with bosonic supports. In other words, computing an expected value of a fermionic operator involves the same number of isometries/disentanglers as in the bosonic MERA, so that local observables can be computed efficiently. 
Finally, we notice that all the previous considerations still apply in lattices where each site is described by a larger vector space $\mathbb{V}$, by decomposing the space $\mathbb{V}_r$ of site $r$ into even and odd parity subspaces $\mathbb{V}_r \cong \mathbb{V}_r^{(0)}\oplus \mathbb{V}_r^{(1)}$, with projectors $P_r^{(0)}$ and $P_{r}^{(1)}$, and defining the $Z_r$ operator as	$Z_r \equiv P_r^{(0)} - P_{r}^{(1)}$. 

In summary, the use of fermionic disentanglers and isometries allows us to maintain the locality of fermionic operators during coarse-graining. To put it in the language of quantum circuits, in which the bosonic MERA was originally formulated: the causal structure of the MERA, consisting of past causal cones with finite 'width' \cite{MERA}, is preserved when replacing bosonic wires and gates with fermionic ones. As a result, both the TTN \cite{TTN} and MERA \cite{2D,algorithm, 2Dsmall} algorithms for two dimensional systems can be extended to fermions. Recall that while a TTN accurately describes small two dimensional lattices, scalable simulations are only possible with the MERA. See Ref. \cite{fMERA} for a thorough technical description of the fermionic TTN and MERA algorithms. 

\begin{figure}[!htb]
\begin{center}
\includegraphics[width=8cm]{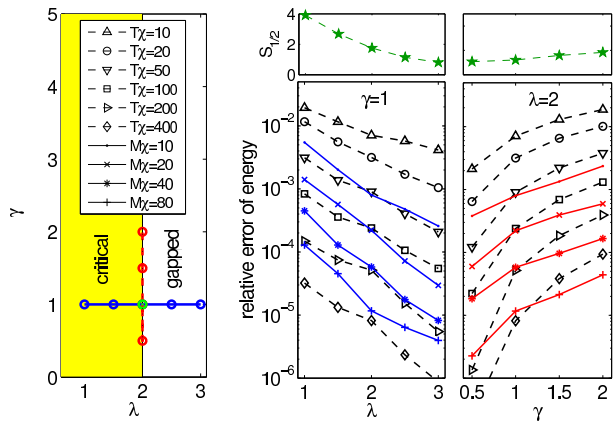}
\caption{
(Color online) Left panel: Phase diagram of the free fermion model (\ref{eq:free}). Right panels: Error in the ground state energy obtained with TTN and MERA simulations of a $6\times 6$ lattice with PBC. The lines correspond to different values of the refinement parameter $\chi$, as indicated in the legend in the left panel. The entanglement entropy $S_{1/2}$ for one half of the lattice is larger for those values of $\lambda$ and $\gamma$ that lead to larger errors.} 
\label{fig:spinless}
\end{center}
\end{figure}

\textbf{Benchmark calculations.---}
To test the validity of the approach, we first consider a system of free spinless fermions with Hamiltonian
\begin{equation}
H_{\mbox{\tiny{free}}} = \sum_{\langle rs \rangle} [c_r^{\dagger}c_s  
+ c_s^{\dagger}c_r - \gamma(c_{r}^{\dagger}c_{s}^{\dagger} +
c_{s}c_{r})] - 2\lambda \sum_r c_{r}^{\dagger} c_r
\label{eq:free}
\end{equation}
on a $6\times 6$ lattice with periodic boundary conditions. This exactly solvable model exhibits a critical (p-wave) superconducting phase for $\gamma>0, 0 < \lambda < 2$, and a gapped superconducting phase for $\gamma>0, \lambda>2$ \cite{Li06, sccomment}. For $\gamma=0$ the pairing potential vanishes and the model corresponds to a free fermion system, i.e. a metal for $0<\lambda<2$ and a band-insulator for $\lambda>2$.
Fig. \ref{fig:spinless} shows the error in the ground state energy as a function of $\gamma$ and $\lambda$, for increasing values of the refinement parameter $\chi$, which is the dimension of the space $\mathbb{V}$ of a coarse-grained site. Both TTN and MERA reproduce several significant digits of the exact solution.\cite{rem:scheme} The entanglement between two halves of the lattice, as measured by the entropy $S_{1/2}\equiv -\tr(\rho\log_2 \rho)$ of the reduced density matrix $\rho$ for half of the lattice, is also plotted. It shows that harder computations (those requiring larger values of $\chi$ in order to achieve a fixed accuracy in the ground state energy) correspond to ground states with more entanglement.  

\begin{figure}[!htb]
\begin{center}
\includegraphics[width=7.3cm]{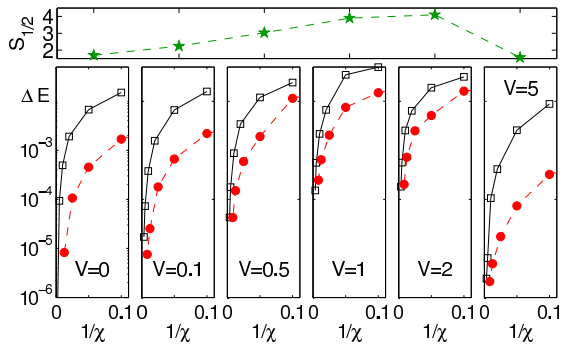}
\caption{
(Color online) Convergence of the ground state energy of interacting spinless fermions (\ref{eq:Hint}) on a $6 \times 6$ lattice with PBC for $\gamma=1, \lambda=2$, and varying interaction strength $V$. The plot shows $\Delta E \equiv E_{\chi}-E_{\chi_{\mbox{\tiny max}}}$, the difference between the energy as a function of $\chi$ (squares for TTN, circles for MERA) and our best MERA result $E_{\chi_{\mbox{\tiny max}}}$, where $\chi_{\mbox{\tiny max}}=120$. Again, the entanglement entropy $S_{1/2}$ shows a strong correlation between ground state entanglement and convergence in $\chi$.} 
\label{fig:ediff}
\end{center}
\end{figure}

\begin{figure}[!htb]
\begin{center}
\includegraphics[width=7.3cm]{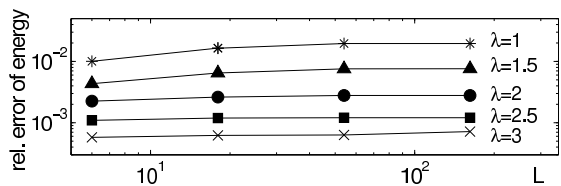}
\caption{
(Color online) The error in the ground state energy as a function of the linear size $L$ of a square lattice, obtained with the fermionic MERA algorithm with $\chi=4$ (non-interacting case, $\gamma=1$), is of the same order of magnitude for small and large systems, even in the critical regime $\lambda<2$. Simulations with interacting fermions (not plotted) show an analogous pattern of energies, but there are no exact results to compare with.
} 
\label{fig:scale}
\end{center}
\end{figure}

Next we add a nearest neighbor repulsion term to $H_{\mbox{\tiny{free}}}$, 
\begin{equation}
	H_{\mbox{\tiny{int}}} = H_{\mbox{\tiny{free}}} + V \sum_{\langle rs \rangle}  c_r^{\dagger}c_r  c_s^{\dagger}c_s,
\label{eq:Hint}
\end{equation}
for which an analytical solution no longer exists. We emphasize that the algorithm does not require any particular modification in order to deal with the interaction. Fig. \ref{fig:ediff} illustrates the convergence of the energy with $\chi$ for different interaction strengths $V$, with $\gamma=1$ and $\lambda=2$. For small and large interaction we observe a similar convergence behavior as in the non-interacting case. For an interaction strength of the order of the hopping amplitude, $V \sim t\equiv1$, the convergence with $\chi$ is slower (the ground state is again more entangled) but about four digits of accuracy seem to still be achieved for large $\chi$. Finally, Fig. \ref{fig:scale} shows the error in the ground state energy for lattices of up to $162\times 162$ sites, demonstrating the scalability of the fermionic MERA algorithm in two spatial dimensions.

\textbf{Discussion.---} 
We have shown that interacting fermionic systems can be addressed within the formalism of entanglement renormalization. Importantly, as with spin systems, the cost of simulations is not determined by the strength of interactions, but by the amount of entanglement in the ground state. While a precise characterization of two-dimensional fermionic systems in terms of ground state entanglement is still missing, our results for interacting fermions are consistent with those obtained in Ref. \cite{FreeFermions} for free fermions and suggest that, broadly speaking, gapped systems are the easiest to simulate. These are followed by gapless systems with a finite number of gapless modes, such as the critical superconducting phase in Fig. \ref{fig:spinless}. Gapless systems with a 1D Fermi surface, the most entangled systems, appear also as the most challenging. 

We envisage that the present approach will help address long standing questions in strongly correlated systems. Presently, a major limitation is due to the scaling $O(\chi^{16}\log L)$ of the computational cost (translation invariant case \cite{2D,fMERA}). While the mild dependence on $L$ ensures scalability, only small values of $\chi$ can be considered. For instance, each $L=162$ simulation of Fig. \ref{fig:scale} with $\chi=4$ already took several days on a 3GHz dual-core desktop PC with 2Gb of RAM \cite{smallL}. A number of strategies are being considered to improve the computational cost, such as alternative coarse-graining schemes, exploitation of internal symmetries (e.g. particle conservation or spin isotropy), use of parallelized code on larger computers, and variational Monte Carlo sampling techniques. Work in progress includes exploring the ground state phase diagram of the Hubbard model.

We thank R. Pfeifer and L. Tagliacozzo for useful discussions, and S. Haas, L. Ding and N. Ali for clarifications concerning the free fermion model \eqref{eq:free}. 
Support from the Australian Research Council (APA, FF0668731, DP0878830) is acknowledged.

\textit{Note added}. Short after this work was made available online, a largely equivalent approach was independently presented by C. Pineda, T. Barthel and J. Eisert in Ref.~\cite{Eisert}.


\begin{thebibliography}{99}
\bibitem{smallL} The cost for $L=6$ is drastically reduced to $O(\chi^4)$.
\end{thebibliography}
\end{document}